\documentclass[12pt]{article}
\usepackage{amssymb}
\usepackage{amsmath}
\usepackage{fullpage}
\usepackage{color}
\usepackage{enumerate}
\usepackage{graphicx}

\newcommand{\mpl}{M_{\mathrm{pl}}}
\newcommand{\ud}{\mathrm{d}}
\newcommand{\pd}{\partial}
\def\D{{\cal D}}
\def\tilD{{\tilde {\cal D}}}
\def\bk{{\mathbf{k}}}
\def\f {{\phi}}
\def\sz{Senatore \& Zaldarriaga}
\def\nablap{\nabla_\parallel}
\def\lag{{\cal L}}
\def\etapl{{\eta_\parallel}}
\def\fnl{f_{\mathrm NL}}
\def\tilK{{\tilde K}}
\begin{document}

\begin{flushright}
\small
MAD-TH-11-07
\date \\
\normalsize
\end{flushright}

\vskip 1cm

\begin{center}
{\large \bf Effective Field Theory and Decoupling in Multi-field Inflation: \\
\vskip 0.2cm
An Illustrative Case Study}

\vspace*{0.5in} {\large Gary Shiu$^{1,2}$ and Jiajun Xu$^1$}
\\[.3in]
{\em $^1$ Department of Physics,
     University of Wisconsin-Madison,
     Madison, WI 53706, USA} \\
     {\em $^2$ Institute for Advanced Study, Hong Kong University of Science \& Technology, 
     Hong Kong}
\\[0.3in]
\end{center}

\vskip 0.5cm

\begin{center}
{\bf
Abstract}
\end{center}
\noindent
We explore the effects of heavy degrees of freedom on the evolution and perturbations of light modes in multifield inflation. We use a simple two-field model as an example to illustrate the
subtleties of integrating out massive fields in a time-dependent background. We show that when adiabaticity is violated due to a sharp turn in field space, the roles of massive and massless field  are interchanged, and furthermore the fields are strongly coupled; thus the system cannot be described by an effective single field action. Further analysis shows that the sharp turn imparts a non Bunch-Davis component in each perturbation mode, leading to oscillatory features in the power spectrum, and a large resonantly enhanced bispectrum. 

\vfill

\newpage

\tableofcontents

\section{Introduction}

Effective field theory (EFT) is a powerful tool for exploring physics whose energy scale exceeds what is currently accessible to us experimentally (and sometimes also theoretically). 
The EFT approach, which emphasizes symmetries, is particularly suited for 
understanding the decoupling of energy scales and the relevant degrees of freedom.
Indeed, this ``bottom-up" perspective, through enumerating the lowest dimension operators compatible with the underlying symmetries, has found wide-ranging  applications, from particle physics to condensed matter systems. One can parametrize our ignorance of short distance physics in a systematic and controlled way before the underlying microscopic theory is fully understood. 

In this regard, inflation is  another natural arena for EFT to find its applications.
While the generic predictions of inflation are in excellent agreement with data, its success 
is
highly sensitive to UV physics.
In particular, a dimension six Planck suppressed operator can give an order one contribution to the slow-roll parameter $\eta$ (which measures the curvature of the potential) and stop inflation, unless there exists  a symmetry (such as shift symmetry) preserved by Planck scale physics to forbid such operator.
The EFT approach thus gives us a recipe to select, among the vast number of inflationary models, those whose UV physics is
compatible with observations.
It would therefore be of interest, both observationally and theoretically,  to formulate a general effective action of inflation, and
indeed some initial forays into the subject can be found in 
\cite{Cheung:2007st,Weinberg:2008hq,Senatore:2010wk}.
In this work, we  follow up on these threads. After laying out our general results  which complement the aforementioned earlier works, we present a worked example which hopefully serve to illustrate some of the subtleties involved.

The procedure of integrating out heavy degrees of freedom in flat space EFT is standard. At energies below the heavy scale $M$, the effects of short distance physics can be summarized by a set of higher dimensional operators, suppressed by  powers of $M$, that are consistent with Lorentz invariance and other underlying symmetries.
This standard procedure, however, becomes more subtle for time-dependent backgrounds.
First of all, time translation as well as 4D Lorentz symmetries are broken, and  many additional operators can arise. For example, ``boundary operators" can be introduced as a way
to set the initial conditions for inflation \cite{Schalm:2004qk,Greene:2004np,Schalm:2004xg,Jackson:2010cw}. Furthermore, even if one can always integrate out the heavy field and describe the classical field dynamics effectively as one single field, one cannot always do so for the quantum perturbations. We will see that non-adiabaticity of a time-dependent background can sometimes cause heavy fields (which naively can be integrated out) to become momentarily light, as well as strong couplings between the light and heavy modes.
Therefore,
 it is worthwhile to revisit some of the standard lores in EFT. Having some solvable examples, especially those that illustrate the subtleties involved,  
 would certainly be welcoming in this regard.

In this work, we illustrate some issues involved in formulating an EFT for inflation
with a two-field model.
The physics and signatures of such model depend as usual on the masses of the fields but also on the classical  trajectory (e.g., its turn rate defined more precisely below) in field space.
When both fields are light and the turn rate is slow (i.e., slow-roll, slow turn), their quantum fluctuations do not freeze after horizon exit but are sourced by each other. Such super-horizon evolutions and their effects on the power spectrum and non-Gaussianities have been well studied \cite{Amendola:2001ni, Gordon:2000hv, Peterson:2010np,Sasaki:1995aw, Vernizzi:2006ve, Meyers:2010rg,GarciaBellido:1995qq,Salopek:1990jq, Rigopoulos:2005xx, Rigopoulos:2005ae}.
When one of the fields is much heavier\footnote{When the mass of the heavy field is comparable to Hubble, we have the quasi-single field scenario \cite{Chen:2009zp}. It is important that the mass of the heavy field is comparable but not much bigger than Hubble so that  its decay is slow enough for the interaction to play a role. The turn rate in this scenario is assumed to be slow.}
 than the Hubble scale during inflation, 
one would naively expect that  this heavy field can be integrated out, resulting in an effective single field model\footnote{It is also worth pointing out that a heavy field can influence the power spectrum if it happens to be excited at the beginning of inflation, and if inflation does not last too long. \cite{Burgess:2002ub}}.
In fact, it was recently argued that the 
result of integrating out the heavy mode can be summarized by an effective sound speed \cite{Tolley:2009fg,Achucarro:2010da}.
While this is true for a slowly-turning trajectory, 
our results show that this effective single field description breaks down when the turn rate is sharp.
We consider a two-field model which is illustrative but at the same time simple enough to be solved exactly.
We further computed the power spectrum as well as the bispectrum for this solvable model, and found that these observables display interesting features that are not captured by a naive effective single field model.
Our work is thus consistent with the recent observation that in  inflationary models with a small sound speed (which, as we shall see, is momentarily the case during the sharp turn), there exists a strong coupling scale below Hubble \cite{Baumann:2011su} which signals the incompleteness of a naive single field description.

This paper is organized as follows. In Section \ref{N-field-EFT}, we discuss the classical dynamics for general multi-field inflation, and present the most general effective action for $N$ canonical scalar field with kinetic mixing to quadratic order of the quantum fluctuations. We then compare our quadratic action to that recently obtained in \cite{Senatore:2010wk} using a Goldstone mode approach.
In Section \ref{Mass Scale}, we revisit the criteria for 
the validity of  EFT based on the classical dynamics of  inflaton \cite{Weinberg:2008hq}. We found that while the mass scale of new physics associated with the tangent direction of the trajectory 
is similarly bounded $M_{\parallel} \geq \sqrt{2 \epsilon} M_P$, the 
mass scale associated with the transverse directions are not subject to this constraint.
In Section \ref{Two Field Example}, we worked out the EFT for perturbations based on a two-field model in detail. We computed the power spectrum and the bispectrum to illustrate that these observables are distinct from what naively expected from an effective single field model, even though
the mass of the heavier field is above Hubble. We end with some discussions and a summary in Section \ref{Summary}.

\section{Comparing Effective Action with Full Action}\label{N-field-EFT}

In this section, we compare the full quadratic level action for $N$ minimally coupled scalar fields in an inflationary background with the effective action based on the Goldstone method by \sz~in Ref.\cite{Cheung:2007st, Senatore:2010wk}. We show that imposing shift symmetry on all the scalar fields and decoupling gravity, while greatly simplifies the analysis for the effective action in Ref.\cite{Senatore:2010wk},
also forbids many contributions that are crucial to account for some interesting multifield dynamics.

\subsection{Classical Background}

We consider a class of multifield inflation models described by the following action 
\begin{equation}
S = \int \ud^4 x \sqrt{-g} \left[ \frac{1}{2} \gamma_{ab} \pd_\mu \phi^a \pd_\nu \phi^b - V(\f^a) \right]
\end{equation}
Here we have $N$ scalar fields $\phi^a$ ($a = 1, 2,\dots, N$). This type of action has been carefully studied in Ref.\cite{GrootNibbelink:2001qt}, and we will follow the formalism therein. 

The homogeneous classical field $\f^a(t)$ follows the equation of motion 
\begin{eqnarray}\label{eom_f1}
\ddot{\f^a}  + \Gamma^a_{bc} \, \dot{\phi}^b \dot{\f}^c + 3H \dot\f^a + \gamma^{ab} \nabla_b V = 0 ~. 
\end{eqnarray}
Introducing the covariant derivative
\begin{equation}
\D_t \dot{\f}^a \equiv \frac{\ud \dot{\f}^a}{\ud t} + \Gamma^a_{bc} \, \dot{\phi}^b \dot{\f}^c ~, \quad \Gamma^a_{bc} = \frac{1}{2} \gamma^{ad}\left( \gamma_{db, c} + \gamma_{dc,b} - \gamma_{bc, d} \right) ~, 
\end{equation}
we can rewrite Eq.(\ref{eom_f1}) in a more concise form
\begin{equation}\label{eom_f2}
\D_t \dot{\f^a} + 3H \dot\f^a + \gamma^{ab} \nabla_b V = 0
\end{equation}

We can define a composite scalar field $\f_0(t)$ through
\begin{eqnarray}
\dot{\f}_0^2 \equiv \gamma_{ab}  \dot{\phi}^a \dot{\phi}^b ~.
\end{eqnarray}
One can show that the equation of motion for $\f_0$ resembles that of a single scalar field 
\begin{equation}\label{eom_f0}
\ddot{\f}_0 + 3H \dot{\f}_0 + \nablap V = 0 ~, 
\end{equation}
with $\nablap V$ the covariant derivative along the tangent direction of the classical inflaton path
\[
\nablap V \equiv \frac{\dot{\f}^a}{\dot{\f}_0} \nabla_a V ~.
\]
The inflationary parameters $\epsilon$ and $\eta$ are defined as usual
\begin{eqnarray}
\epsilon \;\equiv\; -\frac{\dot{H}}{H^2} \;=\; \frac{\dot{\f_0}^2}{2 H^2 \mpl^2} ~, \quad 
\eta \;\equiv\; \frac{\dot{\epsilon}}{H\epsilon} ~.
\end{eqnarray}

\subsection{Kinematic Basis}

The metric $\gamma_{ab}$ can be locally diagonalized by a set of vielbeins $e^a_I$,
\[
e^a_I e^b_J \delta^{IJ} = \gamma^{ab} ~, \quad e^a_I e^b_J \gamma_{ab} = \delta_{IJ} ~. 
\]
In particular, it will be convenient to choose two of the vielbeins pointing along the tangent and normal directions with respect to the classical trajectory, which we denote by
\begin{equation}
e_\zeta^a \equiv \frac{\dot{\phi}^a}{\dot{\f}_0} ~, \quad e_\sigma^a \equiv \frac{\D_t e_\zeta^a}{|\D_t e_\zeta^a|}
\end{equation}

Given $e^a_I e^b_I \gamma_{ab} = 1$, taking covariant time derivative on both sides shows that $e^a$ and $\D_t e^b$ are orthogonal by construction, i.e.,
\begin{equation}
(D_t e^b_I) \, e^a_I \, \gamma_{ab} = 0 ~. 
\end{equation} 
Therefore $e_\zeta$ and $e_\sigma$ are orthogonal by definition. 

The set of vielbeins $\{e_I^a\}$ define a set of complete orthonormal vectors in the field space spanned by $\phi^a$. 
We have denoted two of them by $e_\zeta^a$ and $e_\sigma^a$. The rest are collectively denoted by $e_m^a$. Namely, we have $e_I^a \equiv \{e_\zeta^a, e_\sigma^a, e_m^a\}$ with ($m = 1,2,\dots, N-2$). The $\{a, b, \dots\}$ indices will be lowered and raised by $\gamma^{ab}$, and $\{I, J, \dots\}$ indices will be contracted by $\delta^{IJ}$. 

Another useful parameter is the turn rate $\dot{\theta}$ of the classical trajectory. We define
\begin{equation}
\dot{\theta} = e_{\sigma a} (D_t e_\zeta^a) ~. 
\end{equation}
Using the classical equation of motion Eq.(\ref{eom_f2}) and Eq.(\ref{eom_f0}), we can relate $\dot{\theta}$ to the potential gradient along the $e_\sigma$ direction, i.e.,
\begin{equation} \label{theta_dot}
\dot{\theta} = - \frac{e^a_\sigma \nabla_a V}{\dot{\f}_0} = - \frac{\nabla_\sigma V}{\dot{\f}_0} ~.
\end{equation}
$e^a_\sigma \nabla_a V$ can therefore be understood as the centripetal force to bend the classical trajectory. 

\subsection{The Quadratic Action}

The quadratic action in the spatially flat gauge is given by
\begin{eqnarray}\label{action_q}
S^{(2)} = \frac{1}{2} \int \ud^4 x \; a^3 \left[ \tilD_t Q^I \tilD_t Q^J \delta_{IJ} - \frac{1}{a^2} \pd_iQ^I \pd^i Q^J \delta_{IJ} - m_{IJ} Q^I Q^J \right] ~, 
\end{eqnarray}
Here $Q^I$ is the scalar field perturbation along the kinematic basis $Q^I \equiv e_a^I \delta\f^a$. 

Note that the covariant derivative $\tilD_t$ on $Q^I$ is different from the covariant derivative $D_t$ on the classical field $\dot{\f}^a$. Following Ref.\cite{Achucarro:2010da}, $\tilD_t$ is constructed from the spin connection ${Y^I}_J$, i.e.,
\begin{eqnarray}
\tilD_t Q^I &\equiv& \dot{Q}^I + {Y^I}_J Q^J ~, \\
{Y^I}_J &\equiv& e^I_a D_t e^a_J ~. 
\end{eqnarray}

The mass matrix $m_{IJ} = e^a_I e^b_J m_{ab}$ with $m_{ab}$ given by
\begin{equation}
m_{ab} = M_{ab} - \frac{1}{a^3} \D_t \left[ \frac{a^3\dot{\f}_0^2}{H} e^\zeta_a e^\zeta_b \right]
\end{equation}
Here $M_{ab}$ includes  contributions from both the potential and the curvature of field space
\[
M_{ab} \equiv \nabla_a\nabla_b V + 2\dot{H} {\cal R}_{acdb} \, e^c_\zeta e^d_\zeta ~. 
\]

Switching to conformal time $\ud \tau \equiv \ud t /a$ and changing to canonical variables $v_I \equiv a Q_I$,  we can write the action (\ref{action_q}) in terms of $v_\zeta$, $v_\sigma$ and $v_m$'s. 
\begin{eqnarray}
\lag^{(2)}_{(\zeta)}  &=&  \frac{1}{2} \left( v^{'2}_\zeta - (\pd v_\zeta)^2 + \frac{z''}{z} v_\zeta^2  \right)  ~,
\quad  z \equiv a\dot{\f}_0/H \\
\lag^{(2)}_{(\sigma)} &=&  \frac{1}{2} \left[ v^{'2}_\sigma - (\pd v_\sigma)^2 + \left( \frac{a''}{a} - a^2 M_{\sigma\sigma} + \theta^{'2} - a^2 {Y_{\sigma}}^m Y_{m \sigma} \right) v_\sigma^2 \right]  \label{lag_sig} \\
\lag^{(2)}_{(m)} &=& \frac{1}{2} \left[ v^{'2}_m - (\pd v_m)^2 +  \left( \frac{a''}{a} \delta_{m n} - a^2 M_{m n} + a^2 {Y^I}_m Y_{I n} \right) v_m v_n + 2a Y_{mn} (v_n v_m' - v_m v'_n) \right] \label{lag_m} \nonumber \\ \\
\lag^{(2)}_{(\zeta, \sigma)} &=& \left( - 2\theta' v_\sigma v_\zeta' + 2 \frac{z'}{z}\theta' v_\sigma v_\zeta \right) \label{mix_zs} \\
\lag^{(2)}_{(\sigma, m)} &=& \frac{1}{2} \left( - a^2 M_{\sigma m} + a^2 {Y^I}_\sigma Y_{I m} \right ) v_\sigma v_m 
+ a Y_{\sigma m} (v_m v'_\sigma - v_\sigma v'_m) \label{mix_mn}
\end{eqnarray}

\subsection{The Effective Action for the Goldstone Modes}

Using the St\"{u}ckelberg trick, Ref.\cite{Cheung:2007st, Senatore:2010wk} derived the effective action for the Goldstone mode associated with broken time-diffeomorphism in an inflationary background. To quadratic order, the effective action takes this form: 
\begin{equation}\label{action_pi}
S_{(\pi, \sigma)}^{(2)} = \int \ud^4 x \; a^3 \left[ (2M_2^4 - \mpl^2 \dot{H}) \dot{\pi}^2 +  \mpl^2 \dot{H}\frac{\left(\pd_i\pi\right)^2}{a^2}
+ 2 \tilde{M}_1^{2 I} \dot{\pi} \dot{\sigma}_I + (1 + \tilde{e}_2^I)\dot{\sigma}_I\dot{\sigma}_I 
+ \frac{(\pd \sigma_I)^2}{a^2}
\right]
\end{equation}
Here, the $\pi$ field is the Goldstone mode corresponding to the broken time-diff symmetry, and $\sigma_I$'s are perturbations from the extra light fields. We can establish the following relations
\begin{equation}
v_\zeta = a Q_\zeta = a\dot{\f}_0\pi  ~, \quad v_\sigma = a \sigma_I ~. 
\end{equation}
For the classical background, we also have
\begin{equation}
\mpl^2 \dot{H} = -\dot{\f}_0^2/2 ~, \quad M_2 = 0 ~, \quad \tilde{e}_2^I = 0 ~. 
\end{equation}

We therefore notice that $v_\zeta$ is exactly the canonical $\pi$ field (denoted by $\pi_c$) whose action is
\[
\pi_c^{'2} - (\pd \pi_c)^2 + \frac{(a\dot{\f}_0)''}{a\dot{\f}_0} \pi_c^2
\]
In the limit $\epsilon \to 0$, $H \sim \mathrm{const}$, 
\[
\frac{z''}{z} = \frac{(a\dot{\f}_0)''}{a\dot{\f}_0} ~, 
\]
so the actions for $v_\zeta$ and $\pi_c$ agree. 

Similarly, $v_\sigma$ is the canonical $\sigma$ field $\sigma_c$ whose action is
\[
\sigma_c^{'2} - (\pd \sigma_c)^2 + \frac{a''}{a} \sigma_c^2
\]

Comparing with the action (\ref{lag_sig}), the effective action misses terms such as $\sigma_I \sigma_J$, which can be generated either by a turning trajectory $\theta'\ne 0$, or mass terms in the classical Lagrangian. Comparing with action (\ref{lag_m}) and (\ref{mix_zs}), the effective actions misses terms like $\sigma_I \dot{\sigma}_J$,  $\dot{\pi} \sigma$, $\pi \sigma$. All such terms are not in the effective action because of the shift symmetry imposed on the $\sigma$ fields to keep them light. However, if we 
make no such a priori assumption,
 there will be a lot more terms mixing the $\pi$ field and others fields. As we will see, these mixing terms will generate interesting features when the background inflaton path makes a sharp turn. 

In summary, the Goldstone method does not provide the most general action for multifield perturbations. It is limited in the regime with $\epsilon \to 0$, $\dot{\theta} \to 0$ and all extra fields massless, which 
does not capture many interesting dynamics of multifield inflation especially 
those associated with a
turning trajectory. 

\section{Short Distance Scales}\label{Mass Scale}

In Ref.\cite{Weinberg:2008hq}, Weinberg argued that in any effective field theory of single field inflation, the characteristic mass scale $M$ must be much larger than $\sqrt{2\epsilon} \mpl$. This argument is based on the observation that during inflation, the classical field travels $\Delta \f = \dot\f H^{-1}$ within one e-fold. Using $\epsilon = \dot{\f}^2/(2\mpl^2 H^2)$, we immediately get $\Delta \phi = \sqrt{2\epsilon}\mpl$. Therefore, if $M \lesssim \sqrt{2\epsilon}\mpl$, we expect
that
the effective action which can be expressed in terms of an infinite series expansion of $\f/M$
to receive large, uncontrollable corrections.

The same argument can be applied to multifield inflation, for the mass scales tangent to the classical trajectory. As we have seen in Eq.(\ref{eom_f0}), the composite scalar field $\f_0$ behaves just like a single field, and the effective one field potential is $V(\phi_0) \equiv V(\phi^a(\phi_0))$. Let us denote the mass scale of $V(\f_0)$ by $M_\parallel$, then Weinberg's argument applies. Only if $M_\parallel \gtrsim \sqrt{2\epsilon}\mpl$, we can truncate the potential to finite powers of $\f_0/M_\parallel$.

However, the mass scale associated with the transverse direction easily evades Weinberg's argument, as there is no classical field velocity along transverse directions. However, if the classical trajectory is turning in field space, the classical field will not sit at the minimum of the transverse directions. From Eq.(\ref{theta_dot}), we know that whenever $\dot{\theta} \ne 0$, $\nabla_\sigma V \ne 0$, the field will shift away from the minimum along the transverse directions due to a centrifugal force. 

Assuming that the transverse direction has a potential: 
\[
V_\bot = \frac{1}{2} M_\sigma^2 \, \sigma^2
\]
we can estimate the shift $\Delta \sigma$ to be
\begin{equation}
\Delta\sigma \sim \frac{\nabla_\sigma V}{M_\sigma^2} \sim \frac{\dot{\f}_0 \, \dot{\theta}}{M_\sigma^2} ~.
\end{equation}
We require  $\Delta\sigma$ to
cause little back-reaction on the background. The potential energy lift due to $\Delta\sigma$ is 
\begin{eqnarray}
\Delta V = \frac{1}{2} M_\sigma^2 (\Delta\sigma)^2 \sim \frac{\dot\f_0^2 \, \dot\theta^2}{2 M_\sigma^2}~. 
\end{eqnarray}
These energy comes from changes in the kinetic energy of the inflaton field $\Delta(\dot\f_0^2)$. Due to energy conservation,
\[
3H^2 \mpl^2 = \frac{1}{2}\dot\f_0^2 + V ~,
\]
we conclude that $H$ is not affected by $\Delta \sigma$. 

However, there could potentially be large back-reaction on the $\epsilon$ parameter due to the turn. Since $\epsilon \sim \dot\f_0^2/H^2$ is directly related to the kinetic energy in the inflaton field. If we require that the $\epsilon$ parameter is not changed much during the turn, we need
\begin{equation}\label{bd_msigma}
\Delta V  \ll \frac{1}{2}\dot\f_0^2  \quad \Rightarrow \quad \frac{M_\sigma}{H} \gg \frac{\dot{\theta}}{H}
\end{equation}
This bound on $M_\sigma$ is trivially satisfied if the turn rate is small $\dot\theta/H \ll 1$ and $M_\sigma \gg H$, and can be saturated when the turn is sharp. 

It is interesting to note that the bound on $M_\parallel$ can be written as
\[
M_\parallel \gg \frac{\dot\f_0}{H} ~. 
\]
We see that the linear field velocity $\dot\f_0/H$ set the bound for $M_\parallel$ while the angular velocity $\dot\theta/H$ set the bound for $M_\sigma$. 

For later convenience, we introduce the energy transfer fraction $\beta$
\begin{eqnarray}\label{beta}
\beta \equiv \frac{2 \Delta V}{\dot\f_0^2} = \frac{\dot\theta^2}{M_\sigma^2}
\end{eqnarray}
The bound Eq.(\ref{bd_msigma}) is equivalent to $\beta \ll 1$. Apparently, the energy transfer from the inflaton field to the massive field directly measures how much back-reaction the turn imparts on the classical inflaton trajectory. 

\section{A Two Field Example}\label{Two Field Example}

In the previous section, we have seen that the mass scale along the tangent and orthogonal directions are subject to different bounds. In this section, we will examine the validity of effective single field theory in describing the perturbations of the inflation field. 

In the case of two field model, the quadratic action can be simplified into two parts: the free field action and the quadratic interaction terms. Specifically, we get
\begin{eqnarray}
\lag^{(2)}_{0} &=& \frac{1}{2} \left( v^{'2}_\zeta - (\pd_i v_\zeta)^2 + \frac{z''}{z} v_\zeta^2  \right) + \frac{1}{2} \left[ v^{'2}_\sigma - (\pd_i v_\sigma)^2 + \left( \frac{a''}{a} - a^2 M_{\sigma}^2 + \theta^{'2} \right) v_\sigma^2 \right] 
\label{2field_free} \\
\lag^{(2)}_{\rm int} &=& - 2\theta' v_\sigma v_\zeta' + 2 \frac{z'}{z}\theta' v_\sigma v_\zeta \label{2field_int}
\end{eqnarray}
with $M_{\sigma}^2 = V_{\sigma\sigma} + \epsilon H^2 {\cal R} $, where ${\cal R}$ is the Ricci scalar for the field manifold. 

Introducing the parameters 
\begin{equation}
\eta_\parallel \equiv \frac{V_{\zeta\zeta}}{H^2} ~, \quad \eta_\bot \equiv \frac{M_{\sigma}^2}{H^2} ~,
\quad \varrho \equiv \frac{\dot\theta}{H}, 
\end{equation}
we can expand $z''/z$ as
\begin{eqnarray}
\frac{z''}{z} = a^2 H^2 (2 - \eta_\parallel + \varrho^2 + 5\epsilon + 2\epsilon\eta - 2\epsilon^2) ~,
\end{eqnarray}
and similarly we have 
\begin{eqnarray}
\frac{a''}{a} - a^2 M_{\sigma}^2 + \theta^{'2} = a^2 H^2 (2 - \epsilon - \eta_\bot + \varrho^2 ) ~. 
\end{eqnarray}

Comparing $\lag^{(2)}_0$ with the action of a free massive scalar field $u$ in de-Sitter space, 
\[
\lag = \frac{1}{2} \left( u^{'2} - (\pd u)^2 + a^2 H^2 \left( 2 - \epsilon - \frac{m^2}{H^2} \right) u^2 \right)
\]
we can read off the effective masses for $v_\zeta$ and $v_\sigma$ as
\begin{eqnarray}
m_\zeta^2 &=& H^2 (\eta_\parallel - \varrho^2  - 6\epsilon - 2\epsilon\eta + 2\epsilon^2) ~, \label{mzeta} \\
m_\sigma^2 &=&  H^2 (\eta_\bot - \varrho^2) \label{msigma}
\end{eqnarray}

The physics of the two field system given by $\lag^{(2)}_0 + \lag^{(2)}_{\rm int}$ depends on how the 
parameters $\eta_\parallel$, $\eta_\bot$ and $\varrho$ compare to 1. Generally speaking, we have the following scenarios:

\begin{enumerate}[(a)]

\item \label{scenario1} $\eta_\parallel \ll 1$, $\eta_\bot \ll 1$ and $\varrho \ll 1$: This is two field inflation in the Slow-roll Slow-turn (SRST) regime, which can be solved by treating $\lag^{(2)}_{\rm int}$ as perturbations. The field $v_\zeta$ and $v_\sigma$ will evolve according to the equation of motion derived from $\lag^{(2)}_0$. Due to the interaction between $v_\zeta$ and $v_\sigma$, the field $v_\zeta$ does not freeze after horizon exit. Whenever $\varrho \ne 1$, $v_\sigma$ will source the super-horizon evolution of $v_\zeta$. Such super-horizon evolution can be treated by solving for the transfer functions between $v_\zeta$ and $v_\sigma$ \cite{Amendola:2001ni, Gordon:2000hv, Peterson:2010np} or by using the semi-classical $\delta N$ formalism \cite{Sasaki:1995aw, Vernizzi:2006ve, Meyers:2010rg}, and it has been shown that the two approaches are equivalent \cite{GarciaBellido:1995qq}. One can also study the perturbations using the so called
non-linear long-wavelength approach \cite{Salopek:1990jq, Rigopoulos:2005xx, Rigopoulos:2005ae}. 

\item \label{scenario2} $\eta_\parallel \ll 1$, $\eta_\bot \sim 1$ and $\varrho \ll 1$: This is the quasi-single field scenario studied in Ref.\cite{Chen:2009zp}. Unlike scenario (\ref{scenario1}), one has a massive field $v_\sigma$ with $m_\sigma \sim H$ which is critically damped and will decay after horizon exit. Since $v_\sigma$ is not the inflaton field, it can have large self-interactions, which can mediate interactions among $v_\zeta$ through $\lag^{(2)}_{\rm int}$.  Since $\varrho \ll 1$, we can still treat $\lag^{(2)}_{\rm int}$ perturbatively by introducing transfer vertexes between $v_\zeta$ and $v_\sigma$. It is also important that $\eta_\bot \sim 1$ but not $\gg 1$, so that $v_\sigma$ does not decay too quickly outside the horizon and interaction of $v_\sigma$ can be transferred to $v_\zeta$. 

\item $\eta_\parallel \ll 1$, $\eta_\bot \gg 1$, $\varrho \ne 0$: By conventional wisdom, this scenario should be well described as single field inflation. Even if $\varrho \ne 0$, the massive field $v_\sigma$ quickly decays and settles at the minimum of its potential, so the coupling between $v_\sigma$ and $v_\zeta$ does not seem to play an important role. The light field $v_\zeta$ will undergo the usual horizon exit process as in single field inflation. 

However, Ref.\cite{Tolley:2009fg} and \cite{Achucarro:2010da} showed that by  integrating out the massive mode $v_\sigma$, the resulting effective single field action acquires an effective sound speed $c_s$, which in our notation\footnote{Note that our $\eta_\bot$ is $M^2/H^2$ in Ref.\cite{Achucarro:2010da} and our $\varrho$ is $\eta_\bot$ in Ref.\cite{Achucarro:2010da}} reads, 
\begin{equation}\label{cs_eff}
c_s^{-2} \approx 1 + \frac{4\varrho^2}{\eta_\bot - \varrho^2 - 2 + k^2/(a^2H^2)}  ~.
\end{equation}
Assuming $\eta_\bot \gg 1$ and $k \ll aH$, we get
\[
c_s^{-2} \approx 1 + \frac{4\varrho^2}{\eta_\bot - \varrho^2}  = 1 + \frac{4}{\beta^{-1} - 1} ~,
\]
where $\beta$ is the energy transfer fraction previously defined in Eq.(\ref{beta}). 

We see that $c_s \to 0$ when $\beta \to 1$. In this limit, the turn strongly back-reacts on the inflaton dynamics, with all the inflaton kinetic energy used up to excite the massive field. In the opposite limit $\beta \ll 1$, back-reaction is negligible and $c_s^{-1} \sim 1 + 4\beta \approx 1$. The behavior of $c_s$ as a function of $\beta$ is shown in Fig. \ref{fig_cs}. 

\begin{figure}[h]
\begin{center}
\includegraphics[width=9cm]{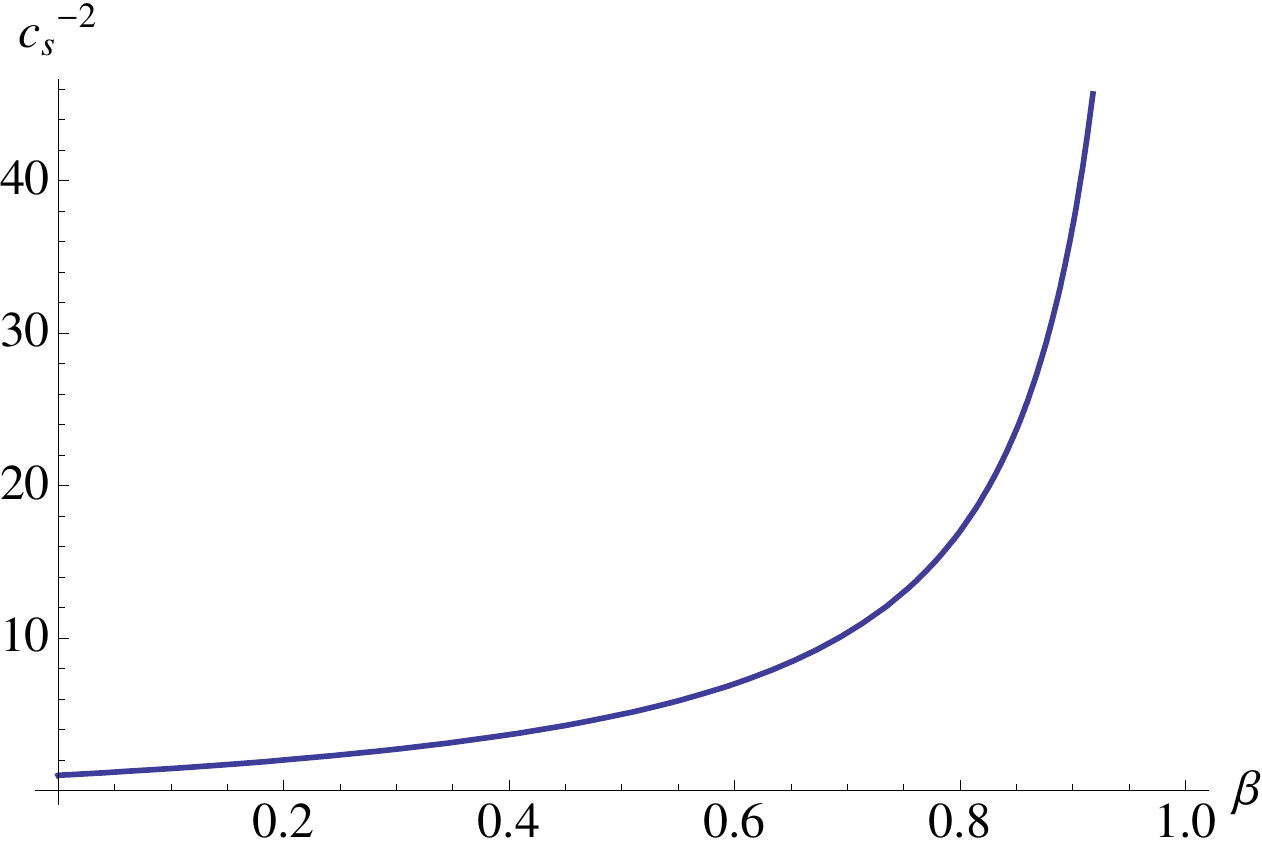}
\caption{\label{fig_cs} \small The effective sound speed as a function of the energy transfer fraction $\beta$. When $\beta \to 1$, $c_s^{-2} \gg 1$; however, back-reaction and coupling are strong in this limit. If $\beta \ll 1$, back-reaction is negligible, but $c_s$ cannot deviate much below 1. }
\end{center}
\end{figure}

We see that when the turn rate $\varrho$ starts to saturate the bound $\varrho^2 \le \eta_\bot$. Two effects happen at the same time. First, $m_\zeta^2/H^2 \sim \varrho^2 \gg 1$ and $m_\sigma^2/H^2 \sim \eta_\bot -\varrho^2 \to 0$, the original massive and massless modes interchanges their role. Physically, this is because a sharp turn causes a quick rotation in the field space, interchanging the original massive and massless directions. The mass hierarchy momentarily vanished during the sharp turn, so we should not integrate out either field. Second, when $\varrho \sim \eta_\bot \gg 1$, the two modes $v_\zeta$ and $v_\sigma$ becomes strongly coupled, and one should solve the full quadratic action $\lag^{(2)}_0 + \lag^{(2)}_{\rm int}$ to obtain the mode functions \cite{Cremonini:2010ua}. 

The fact that an extremely small effective sound speed $c_s$ corresponds to strong coupling and back-reaction on the inflaton field is consistent with the study in Ref.\cite{Leblond:2008gg}, where it was shown that the single field inflaton action is not under perturbative control when $c_s$ becomes extremely small. To deal with the strongly coupled inflaton field, Ref.\cite{Baumann:2011su} suggested a weakly coupled UV completion by introducing a second massive field into the single field effective action. Their perspective was to start from the low energy effective theory and analyze when the effective theory breaks down. Here using the UV complete two field action, we  have clarified from a top down point of view, how the effective single field description breaks down when the classical field trajectory makes a sharp turn. 

In fact, the UV completed action in Ref.\cite{Baumann:2011su} can be casted into the two field action presented here. By going to the low energy limit $k/aH \ll \varrho^2$, one can neglect the usual kinetic terms in $\lag^{(2)}_0$ and the term $v_\zeta' v_\sigma$ in $\lag^{(2)}_{\rm int}$ becomes the non-relativistic kinetic term of the system. This is equivalent to studying the super-horizon evolution of the two field model. 

Following Ref.\cite{Baumann:2011su}, the relevant terms in the action are now
\begin{eqnarray}
{\tilde \lag}^{(2)} &=& - 2\theta' v_\sigma v_\zeta' - \frac{1}{2} (\pd_i v_\zeta)^2 - \frac{1}{2} (\pd_i v_\sigma)^2 \nonumber \\
&& + a^2H^2 \left (2-\epsilon - \frac{m_\zeta^2}{H^2} \right) v_\zeta^2 + a^2 H^2 \left(2-\epsilon - \frac{m_\sigma^2}{H^2} \right) + 2 \frac{z'}{z}\theta' v_\sigma v_\zeta ~. 
\end{eqnarray}
Note that Ref.\cite{Baumann:2011su} followed the Goldstone approach in Ref.\cite{Senatore:2010wk} so that in the effective multi-field action, the $m_\zeta$ term and the $v_\zeta v_\sigma$ term were not allowed by shift symmetry. They also took the decoupling limit $\epsilon \to 0$. Here we are not constrained by shift symmetry and we do not decouple gravity, so all those terms are allowed. 

\end{enumerate}

\subsection{Sharp Turning and the 2-Point Function}\label{2pt}

In this section, we will solve for the mode functions of the strongly coupled two field system under the sharp turn approximation. By sharp turn, we mean that the turn rate $\varrho$ is momentarily large. The time scale of changes in $\varrho$ is much shorter than the oscillation time scale of $v_\zeta$ and $v_\sigma$, and also much less than one e-fold. 

A momentarily large $\varrho$ can be caused either by a sharp feature in the scalar potential or by momentarily large kinetic mixing. When $\varrho \gg 1$, the adiabatic and isocurvature modes are strongly coupled and solving for the mode equations requires the full Lagrangian $\lag^{(2)}_0 + \lag^{(2)}_{\rm int}$. However, a large $\varrho$ may spoil the scale invariance of the power spectrum, as we have seen that $\varrho^2$ contributes to the effective mass of $v_\zeta$. In fact, from the background equation (\ref{eom_f0}), we get
\begin{equation}
\varrho^2 - \eta_\parallel = \frac{\dddot{\phi_0}}{H^2 \dot{\phi}_0} + \frac{3}{2}\eta - 6\epsilon
\end{equation}
If we require slow-roll dynamics along the tangent direction of the trajectory, $\varrho^2 - \eta_\parallel$ has to remain small, and $v_\zeta$ remains a massless field. However, just as in single field inflation models, one can momentarily violate slow-roll conditions. Here, we could have $\dddot{\phi_0}/(H^2\dot{\phi}_0)$ momentarily large due to an sudden acceleration along the tangent direction, which will allow $\varrho^2 \gg \etapl$ momentarily.

When $\varrho$ is momentarily much greater than 1, two effects happen at the same time. First, there will be sudden changes in the mass parameters $m_\zeta$ and $m_\sigma$. From Eq.(\ref{mzeta}) and Eq.(\ref{msigma}), we have
\begin{eqnarray}
m_\zeta^2 &\approx& -\frac{\dddot{\phi_0}}{\dot{\phi_0}} \\
m_\sigma^2 &\approx& H^2 (\eta_\bot - \varrho^2)
\end{eqnarray}
where we have ignored subleading terms ${\cal O}(\epsilon, \eta, \epsilon^2)$, and keep only the terms that dominate at the time of the sharp turn. The momentarily large $m_\zeta$ is analogous to that induced by sharp features in single field inflation. It will generated sinosodial running features in the power spectrum and bispectrum as studied by Ref.\cite{Chen:2006xjb}. In this sense, as far as the $v_\zeta$ alone is concerned, a momentarily large $\varrho$ can be mimicked by sharp features in the single field potential. A second effect of large $\varrho$ is the strong coupling between $v_\zeta$ and $v_\sigma$. Physically, the perturbations along the massive direction get projected into the inflaton direction due to the sharp turn. This effect is multi-field in nature. 

In reality, the background dynamics due to a sharp turn can be very complicated. Generically, one expects that a fraction $\beta$ of the inflaton kinetic energy gets transferred into the potential energy of the massive field, driving the massive field away from the bottom of its potential, therefore providing the centripetal force for the sharp turn. After the sharp turn, the potential energy in the massive field will be converted back into the kinetic energy, which will cause classical oscillations in the heavy field. As studied in Ref.\cite{Chen:2011zf}, such oscillation can trigger resonant enhancement \cite{Chen:2008wn} of non-Gaussianity for the massless $v_\zeta$ field. 

\begin{figure}[h]
\begin{center}
\includegraphics[width=7cm]{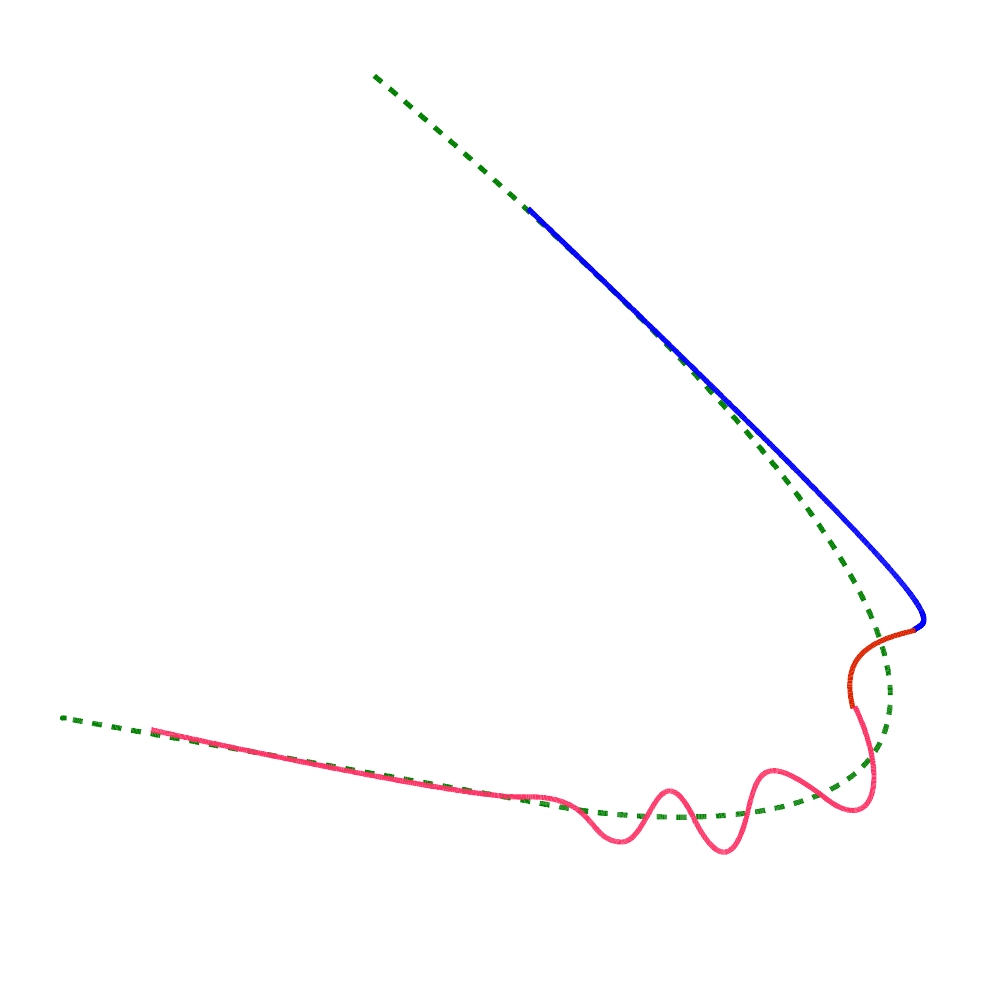}
\caption{\label{fig_turn} \small Illustration of a sharp turn in  field space. The green dashed line represents the massless field (inflaton) direction, and the massive field is orthogonal to the green dashed line. Along the blue part of the trajectory, the kinetic energy of the inflaton is gradually transformed into the potential energy of the massive field. At the end of the blue line, the potential enegry starts to convert back to the kinetic energy, and caused subsequent oscillations of the massive field along the transverse direction (red curve). The perturbations in the massive field are projected into the inflaton direction at the interface between the blue and red curve, and this is the effect we focus on in this paper. The turning angle shown in this figure is made very large to illustrate the excitation of massive modes. In reality, a much smaller turning angle is sufficient, and we will show that $\Delta \theta \lesssim 0.1$ from constraints on the power spectrum. }
\end{center}
\end{figure}

In this paper, we will focus on the scenario when the coupling between $v_\zeta$ and $v_\sigma$ provides the dominant effect at the time of the turn, while the sudden change in $m_\zeta$ is sub-leading. We now identify the parameter region of this scenario. As we have seen, during the turn, a fraction of the inflaton kinetic energy becomes the potential energy in the massive field, this can be modeled as a step function in $\epsilon$,
\[
\epsilon \to \epsilon_0 ( 1 + \beta \, \Theta(\tau - \tau_0)) ~.
\]
The corresponding change in $z = a\sqrt{2\epsilon}$ is given by
\[
z \to z_0 \left(1 + \frac{\beta}{2} \, \Theta(\tau - \tau_0) \right) ~,
\]
Therefore,
\begin{equation}
\frac{z''}{z} = \frac{\beta}{2} \frac{\ud}{\ud \tau} \delta(\tau-\tau_0) + aH \beta \delta (\tau - \tau_0)  + \dots ~. 
\end{equation} 
We see that the singular terms in $z''/z$ or $m_\zeta$ is of magnitude $\beta$. 

On the other hand, energy conservation gives
\begin{eqnarray}
\frac{1}{2} \beta \dot\f_0^2 &=& \Delta V \;=\; \frac{1}{2} (1-\beta) \dot\f_0^2 \, \frac{\dot\theta^2}{M_\sigma^2} \\
\frac{\dot\theta^2}{H^2} &=& \frac{\beta}{1-\beta} \frac{M_\sigma^2}{H^2} \; \approx \; \beta \frac{M_\sigma^2}{H^2}
\end{eqnarray}
For our purpose, we choose
\begin{equation}
\beta \ll 1 ~, \quad M_\sigma \gg H ~, \quad \beta \frac{M_\sigma^2}{H^2} \gg 1 ~,
\end{equation}
so that the features in $z''/z$ is sub-leading, and the leading effects is generated by the strong coupling between $v_\zeta$ and $v_\sigma$. The effect we discuss here precedes the resonant effect discussed in Ref.\cite{Chen:2011zf}.

In Ref.\cite{Achucarro:2010da}, the authors studied the effect of sharp turn due to kinetic mixing. However, the effective single field theory approach in their work requires that the heavy field lies in the adiabatic minimum of its potential. We have seen that this assumption generically does not hold when a sharp turn happens, especially when it excites massive field oscillations. Therefore we will perform a full two field analysis in dealing with the sharp turn. 

In Ref.\cite{Cremonini:2010ua}, the strongly coupled two field system was studied numerically. Here we show that if the coupling $\varrho$ is momentary large, i.e. the time scale of change in $\varrho$ is much shorter than the oscillation time scale of $v_\zeta$ and $v_\sigma$, we can obtain the mode functions analytically. 

Let us start by writing down the mode equations for $v_\zeta$ and $v_\sigma$ derived from the full quadratic Lagrangian $\lag^{(2)}_0 + \lag^{(2)}_{\rm int}$. Introducing a new variable $x \equiv k\tau$, the equations can be written as
\begin{eqnarray}
\frac{\ud^2 v_\zeta}{\ud x^2} + \left(1 - \frac{2}{x^2} \right) v_\zeta - \frac{2\varrho}{x^2} v_\sigma + \frac{\ud}{\ud x}\left(\frac{2\varrho}{x} v_\sigma\right) &=& 0 \\
\frac{\ud^2 v_\sigma}{\ud x^2} + \left(1 - \frac{2-\eta_\bot +\varrho^2}{x^2} \right) v_\sigma - \frac{4\varrho}{x^2} v_\zeta - \frac{\ud}{\ud x}\left(\frac{2\varrho}{x} v_\zeta\right) &=& 0 
\end{eqnarray}

We approximate the momentarily large turn rate as a delta function\footnote{A delta function energy transfer between two fields with exponential potentials has been studied numerically in Ref.\cite{Ashoorioon:2008qr}.}, 
\begin{equation}
\varrho = \frac{\dot{\theta}}{H} = \frac{\Delta \theta}{H} \, \delta(t - t_0) = \Delta \theta\, x_0 \, \delta(x-x_0) ~.
\end{equation}
Matching the mode fucntions before and after the sharp turn, we require
\begin{eqnarray}
v_\zeta\Big|^{x_0+}_{x_0-} &=& -2 \Delta\theta \, v_\sigma \Big|_{x_0-} ~, \\
\frac{\ud v_\zeta}{\ud x} \Big|^{x_0+}_{x_0-} &=& \frac{2\Delta\theta}{x_0} v_\sigma\Big|_{x_0-} ~. 
\end{eqnarray} 

$v_\zeta$ is a massless scalar field in de-Sitter space before and after the turn, so we have
\begin{eqnarray}
v_\zeta (x < x_0) &=& v^+(k, \tau) ~, \\
v_\zeta (x > x_0) &=& C_1 v^+(k, \tau) + C_2 v^- (k, \tau) ~, \\
v^{\pm} (k, \tau) &=& \frac{-1}{\sqrt{2k}} e^{\mp i x} \left( \frac{1}{x}  \pm i \right) ~.
\end{eqnarray}

$C_1$ and $C_2$ can be solved by matching the boundary conditions at $x_0$. 
\begin{eqnarray}
C_1 &=& 1 + i \Delta\theta \, e^{i x_0} \sqrt{2k}\, v_\sigma(x_0) ~, \\
C_2 &=& - i \Delta\theta \, e^{-i x_0} \sqrt{2k}\, v_\sigma(x_0) ~. \label{c2}
\end{eqnarray}

The power spectrum is given by
\begin{equation}
P_\zeta = \frac{k^3}{2\pi^2} \left|\frac{v_\zeta}{a\sqrt{2\epsilon}}\right|^2_{x\to 0} = \frac{H^2}{8\pi^2\epsilon} |C_1+C_2|^2
\end{equation}
The factor $|C_1 + C_2|^2$ encodes all the features in the 2-point function generated by the sharp turn. The full expression of $|C_1 + C_2|^2$ involves the value of the massive mode function at the time of the sharp turn $v_\sigma(x_0)$. 

In the asymptotic limit $x_0 \ll -1$, i.e. for modes inside the horizon at the time of the sharp turn, we have 
\[
\sqrt{2k}\, v_\sigma(x_0) \sim e^{-ix_0} ~. 
\]
Therefore, 
\begin{eqnarray}
|C_1 + C_2|^2 \approx 1 + 2\Delta\theta \sin\left(\frac{2 k}{k_0} \right) ~, \quad k_0 \equiv \frac{-1}{\tau_0} ~, 
\quad k/k_0 \gg 1 ~,
\end{eqnarray}

In limit $x_0 \to 0$, i.e. for modes outside the horizon at the time of the turn, we have
\[
\sqrt{2k}\, v_\sigma(x_0) \sim \sqrt{-\pi x_0} \, \exp\left(-\frac{\pi}{2} \frac{m}{H} \right) \frac{1}{\Gamma(i{\tilde\nu} + 1)} \left( - \frac{x_0}{2} \right)^{i {\tilde\nu}} ~, \quad {\tilde \nu} \equiv \sqrt{\eta_\bot - \frac{9}{4}}
\]
The massive field decays outside the horizon due to the factor $\sqrt{-x_0}$, so as expected 
\begin{equation}
|C_1 + C_2| \to 1 ~, \quad x_0 \to 0 ~. 
\end{equation}
The projection of massive field perturbations into the inflaton direction generates sinusodial ripples on the power spectrum. The level of such oscillations depends on the turning angle $\Delta\theta$. It is interesting that even if the sharp turn happens before 60 efolds from the end of inflation, it may still leave some features in the power spectrum. The chances of detecting such small oscillations in the power spectrum may be small, so we will have to look at the bi-spectrum to search for correlated signatures.

\subsection{The 3-Point Function}

We have seen that when the energy transfer is small $\beta \ll 1$, $c_s \approx 1$. So from the effective single field point of view, the level of equilateral non-Gaussianity is very small. When $\beta \to 1$, $c_s \ll 1$ superficially, but the strong coupling and back-reaction renders the effective single field description invalid. In this section, we will identify the features in 3-point function associated with the sharp turn. 

We perform the computation using the standard in-in formalism \cite{Maldacena:2002vr}. 
\[
\langle \zeta^3 \rangle = -i \int \ud t \langle [\zeta^3, H_I(t) ] \rangle 
\]
with 
\[
\zeta(\bk, \tau) \equiv \frac{v_\zeta(\bk, \tau)}{a\sqrt{2\epsilon}} = u(\bk, \tau) a_\bk + u^*(-\bk, \tau) a_{-\bk}^\dagger. 
\]
\[
u(\bk, \tau) = \frac{iH}{\sqrt{4\epsilon k^3}}(1+ik\tau) e^{-ik\tau}
\]
\[
[a_\bk, a^\dagger_{\bk'}] = (2\pi)^3 \delta^{(3)} (\bk + \bk')
\]
For example, we can consider the 3-point vertex 
\[
H_I = - \int \ud x^3  a^3 \epsilon^2 \zeta \zeta'^2 ~,
\]
which gives
\begin{eqnarray}
\langle \zeta^3 \rangle &=& \left[ i (u_{\bk_1} u_{\bk_2} u_{\bk_3})|_{\tau=0}
 \int_{-\infty}^0 \ud\tau \, a^2 \epsilon^2\, u^*_{\bk_1}(\tau) \frac{u^*_{\bk_2}(\tau)}{\ud \tau} \frac{u^*_{\bk_3}(\tau)}{\ud \tau}
(2\pi)^3 \delta^{(3)}\left(\sum \bk_i\right) + \mathrm{perm}. \right] + c.c. \nonumber \\ \label{3pt_int}
\end{eqnarray}
Using the ansatz for the 3-point function
\begin{equation}
\langle \zeta^3 \rangle = \fnl(k_1, k_2, k_3) \, \frac{P_\zeta^2}{k_1^2 k_2^2 k_3^2} \, (2\pi)^7 \, \delta^{(3)}\left(\sum \bk_i\right) ~,
\end{equation} 
we find that 
\begin{eqnarray}
\langle \zeta^3 \rangle \sim \frac{H^4}{\epsilon} ~, \quad \fnl \sim {\cal O}(\epsilon)  ~. 
\end{eqnarray}
This is the standard result for slow-roll inflation. 

Here we will focus on the effect of the non Bunch-Davis component in the mode function. The major difference in the computation is that starting at some time $\tau_0$, we will have a non Bunch-Davis component, namely the $C_2$ component in Eq.(\ref{c2}). The corrections to the 3-point function can be obtained by first flipping the sign of one of the $k_i$ in Eq.(\ref{3pt_int}) and then adding an overall factor $|C_2|^2$ in front of the integral. The integral will start from a finite $\tau_0$ instead of $-\infty$, but this will not affect an order of magnitude estimation on $\fnl$. In the end, we get
\begin{eqnarray}
\langle \zeta^3 \rangle \sim \frac{H^4}{\epsilon} |C_2^2| ~, \quad \fnl^{\textrm{non BD}} \sim {\cal O}(\epsilon) |C_2^2| ~. 
\end{eqnarray}
This effect of non Bunch-Davis component on $\fnl$ was previously studied in Ref.\cite{Chen:2006nt,Holman:2007na, Meerburg:2009ys}. Here we have provided one microscopic origin of such a non Bunch-Davis component -- a sharp turn in multifield inflation. 

However, observationally, $\fnl \sim \epsilon |C_2|^2$ is very hard to detect. Since the oscillation in the power spectrum is controlled by $|C_2|$. Based on current observational data, assume that the amount of oscillation in the un-binned data is about $\lesssim 10\%$ \cite{Komatsu:2008hk} and take $\epsilon \sim 0.01$, we get negligibly small $\fnl$ on the order of $10^{-4}$.

When the scalar field makes a sharp turn, it is very likely that the massive field will oscillate after the turn, as we have discussed in Sec.\ref{2pt}. We ignored such effects in computing the 2-pt function, however, such oscillations will be important in amplifying the 3-pt function, through the resonant mechanism discussed in Ref.\cite{Chen:2008wn, Chen:2011zf}. The vertex responsible for resonant non-Gaussianity is 
\begin{equation}\label{res_hi}
H_I = -\int \ud \tau \ud x^3 \frac{1}{2} a^2 \epsilon \dot{\eta} \, \zeta^2 \zeta' ~. 
\end{equation}
In usual slow roll limit, such term is sub-leading as $\epsilon\dot{\eta} \sim {\cal O} (\epsilon^3)$. However, in a time dependent oscillating background with
\begin{eqnarray}
\dot{\eta} = (\dot{\eta})_0 + (\dot{\eta})_A \sin \omega t ~, \quad \omega \gg H ~. 
\end{eqnarray} 
we will have resonance enhanced non-Gaussianity
\begin{eqnarray}
\fnl^{\mathrm res} = \frac{(\dot{\eta})_A}{H} \left( \frac{\omega}{H} \right)^{1/2} \frac{\sqrt{\pi}}{8\sqrt{2}} \sin \left( \frac{\omega}{H} \ln K + \phi \right) ~, \quad K = k_1 + k_2 + k_3 ~. 
\end{eqnarray}
Here $\phi$ is a phase independent of $k$. Note that the enhancement of $\fnl$ comes from $\omega \gg H$. The positive power of $\omega/H$ appears counter-intuitive, as one would expect the effect suppressed by powers of $H/M$. However, one should note that the resonance effect is purely due to the time dependence in the coupling constant. There is no suppression of $H/M$ in the interaction term (\ref{res_hi}). In fact, $\sqrt{\omega/H}$ counts the number of resonance periods from which the contribution to $\fnl$ is dominant. 

The oscillation in $\dot\eta$ is sourced by the massive field oscillations, which according to Ref.\cite{Chen:2011zf} give
\[
\frac{(\dot\eta)_A}{H} = \beta \frac{M_\sigma^2}{H^2} ~, \quad \omega = 2M_\sigma ~.  
\]

With a sharp turn introducing the non Bunch-Davis component, we replace $K$ by $\tilK_i \equiv K - 2k_i$ and multiply $\fnl$ by $|C_2|^2$. Therefore, for $k_i > k_0$, the asymptotic form of $f_{\mathrm NL}$ is 
\begin{eqnarray}
f_{\mathrm NL}^{\mathrm{res}}|_{\textrm{non BD}} \sim 
\frac{\sqrt{\pi}}{8}\, \beta\, \left( \frac{M_\sigma}{H} \right)^{5/2} (\Delta \theta)^2
\sin \left( \frac{2M_\sigma}{H} \ln \tilK_1 + \phi \right)  + \mathrm{perm} ~.
\end{eqnarray}

Resonant enhancement of 3-point function has been considered in various scenarios for Bunch-Davis state \cite{Chen:2008wn, Chen:2011zf, Flauger:2009ab}\footnote{The effects of non Bunch-Davis initial states on even the trispectra have been studied in \cite{Chen:2009bc}.}. Generically, they give signatures running in $k$ space according to
\[
\fnl^{\mathrm{res}}|_{\textrm{BD}} \sim \sin \left( \frac{2M_\sigma}{H} \ln K + \mathrm{phase} \right)
\]
For resonance in the non Bunch-Davis components, the running is along the $\tilK_i$ directions, 
\[
\fnl^{\mathrm{res}}|_{\textrm{non BD}} \sim \sin \left( \frac{2M_\sigma}{H} \ln \tilK_1 + \mathrm{phase} \right) + \mathrm{perm} ~.
\]
Comparing the amplitude of the signature, the effect from non Bunch Davis components is suppressed by a factor of $(\Delta \theta)^2$. Suppose from the power spectrum, we estimate that $\Delta \theta \lesssim 0.1$, and let us further assume that $\beta \sim 0.01$, $M_\sigma /H \sim 100$, then we get
\begin{equation}
\fnl^{\mathrm{res}}|_{\textrm{non BD}}  \sim 10 \; \sin \left( \frac{2M_\sigma}{H} \ln \tilK_1 + \mathrm{phase} \right) + \mathrm{perm} ~. \label{mag1}
\end{equation}

Ref.\cite{Chen:2011zf} also studied  resonance enhancement in two field inflation models. The oscillations of the background massive field provide a periodic time dependent background that triggers the resonant effect in the 3-point function. The main difference here is that we focus on the enhancement of the non Bunch-Davis component. 

The resonant enhancement of $\fnl |_{\mathrm{non BD}}$ has also been discussed in Ref.\cite{Chen:2010bka} in the context of single field inflation, where the non Bunch-Davis component is generated by the oscillation of inflaton field. Here we generate the non Bunch-Davis component through a sharp turn that couples the massive field perturbation and the inflaton perturbation. 

Last, we comment that if the inflaton action is a general $p(X)$ where $X \equiv \gamma_{ab} \pd_\mu \phi^a \pd_\nu \phi^b$, the effect we consider here will further be enhanced by a small sound speed $c_s \ll 1$. The relevant interaction Hamiltonian is 
\begin{equation}
H_I = -\int \ud \tau \ud x^3 \frac{a\epsilon}{Hc_s^2}\left(\frac{1}{c_s^2} - 1 - \frac{2\lambda}{\Sigma}\right) \zeta'^3 ~,
\end{equation} 
which leads to 
\begin{eqnarray}
f_{\mathrm NL}^{\mathrm{res}}|_{\textrm{non BD}} \sim \left(\frac{1}{c_s^2} - 1 - \frac{2\lambda}{\Sigma}\right) 
\left(\frac{c_s^2}{\epsilon}\right)_0 \left( \frac{\omega}{H} \right)^{5/2} \beta \left(\Delta \theta\right)^2
\end{eqnarray}
Comparing to the result in Eq.(\ref{mag1}), we expect the signature still has a $\sin(\ln \tilK_i)$ running with its amplitude a factor of $c_s^{-2}$ larger than Eq.(\ref{mag1}).

\section{Summary and Discussions}\label{Summary}

In this paper, we investigated the subtleties involved in integrating out massive degrees of freedom in the inflationary background. We show that a generic multifield inflation action has many interesting terms that are forbidden by imposing shift symmetry on all the fields or by ignoring gravity. Such terms leads to interesting physics, such as turning trajectories for multifield inflation. 

Using a two field system as illustration, we show that the separation of degrees of freedom based on  mass scales could break down in a time dependent background. Especially, we show that during a sharp turn of the two field system, the massive field and massless field interchanges their role, and the effective single field description breaks down. If one naively integrates out the massive field, and defines an effective sound speed for the resulting single field system, one would see that the effective sound speed goes to zero when the turn rate saturates the upper bound $\dot\theta \lesssim M_\sigma$.  When $\dot\theta \sim M_\sigma$, the energy transfer from the massless field to the massive field is maximum, and the massive field is maximally excited. This is an explicit example showing that when the effective sound speed is small, the system cannot be described by a single degree of freedom. In fact, from a top down point of view, we have provided an example of UV completion to the effective single field action with small sound speed. Our example is consistent with previous observations in Ref.\cite{Leblond:2008gg, Baumann:2011su}.

Furthermore, the turn rate $\varrho$ also serves as the coupling between the massive and massless field. When $\varrho \gg 1$, the system becomes strongly coupled and one should solve the full system to capture the dynamics of the mode function. In particular, we investigated the limiting case of a sharp turn modeled by a delta function turn rate. We show that generically, the sharp turn will impart a non-Bunch-Davis component on each mode function, which leads to sinusoidal running in the power spectrum. Generically, we can summarize the oscillatory feature as
\begin{equation}
\frac{\Delta P_\zeta}{P_{\zeta}} = 2\Delta\theta \sin(2k/ k_0) ~. 
\end{equation}
with $\Delta\theta$ the amplitude of non-Bunch-Davis component in the mode function. 

In most cases, the non-Bunch Davis component has non-observable effects on the 3-pt function. However, just as a periodic background can resonantly amplify the 3-pt function for Bunch-Davis component, it does so to the non-Bunch-Davis component as well. As a rule of thumb, we expect
\begin{eqnarray}
\fnl^{\mathrm{non-BD}} \sim (\fnl^{\mathrm{BD}})_A \, |\Delta\theta|^2 \sin (\ln \tilK_i + \mathrm{phase}) + \mathrm{perm.} ~. 
\end{eqnarray}
with $(\fnl^{\mathrm{BD}})_A$ being the amplitude of the signal for the Bunch-Davis components. Note that even if the non-Bunch-Davis component gives the amplitude with a factor of $\Delta\theta^2$ smaller, the signal has a different sinusoidal running from the Bunch-Davis result, i.e. $\sin (\ln \tilK_i)$ instead of $\sin (\ln K)$. So if $(\fnl^{\mathrm{BD}})_A$ is large enough to be detected, a correlated signal of $\fnl^{\mathrm{non-BD}}$, together with oscillations in the power spectrum,
may give us a strong hint that adiabaticity is  violated momentarily during inflation.

\section*{Acknowledgment}
  We thank Amjad Ashoorioon,
  Xingang Chen, Jinn-Ouk Gong, Lam Hui, Louis Leblond, Joel Meyers, Koenraad Schalm,
  Leonardo Senatore, Sarah Shandera, Scott Watson and Matias Zaldarriaga for discussions.
    GS would like to thank the Hong Kong Institute for Advanced Study for their hospitality while this work was completed.
GS is supported in part by a DOE grant under contract DE-FG-02-95ER40896, and a Cottrell Scholar Award from Research Corporation.

\end{document}